\documentclass[%
reprint,
superscriptaddress,
showpacs,preprintnumbers,
amsmath,amssymb,
aps,
onecolumn
]{revtex4-2}

\usepackage{graphicx}
\usepackage{dcolumn}
\usepackage{bm}
\usepackage{xcolor}
\usepackage{braket}
\usepackage{url}
\usepackage{here}

\newcommand{\img}{\mathrm{i}}
\newcommand{\br}{{\bf r}}
\newcommand{\bk}{{\bf k}}

\newcommand{\nm}{\nonumber}

\begin{document}

\preprint{APS/123-QED}

\title{
Fast simulation for multi-photon, atomic-ensemble quantum model of linear optical systems addressing the curse of dimensionality
}

\author{Junpei Oba}
\email{junpei-oba@mosk.tytlabs.co.jp}
\affiliation{Toyota Central R\&D Labs., Inc., 41-1, Yokomichi, Nagakute, Aichi 480-1192, Japan.}
\author{Seiji Kajita}
\email{fine-controller@mosk.tytlabs.co.jp}
\affiliation{Toyota Central R\&D Labs., Inc., 41-1, Yokomichi, Nagakute, Aichi 480-1192, Japan.}
\author{Akihito Soeda}
\email{soeda@nii.ac.jp}
\affiliation{Principles of Informatics Research Division, National Institute of Informatics, 2-1-2 Hitotsubashi, Chiyoda-ku, Tokyo 101-8430, Japan}

\date{\today}

\begin{abstract}
Photons are elementary particles of light in quantum mechanics, whose dynamics can be difficult to gain detailed insights, especially in complex systems.
Simulation is a promising tool to resolve this issue, but it must address the curse of dimensionality, namely, that the number of bases increases exponentially in the number of photons.
Here we mitigate this dimensionality scaling by focusing on optical systems composed of linear optical objects, modeled as an ensemble of two-level atoms. We decompose the time evolutionary operator on multiple photons into a group of time evolution operators acting on a single photon. Since the dimension of a single-photon time evolution operator is exponentially smaller than that of a multi-photon one in the number of photons, the decomposition enables the multi-photon simulations to be performed at a much lower computational cost.
We apply this method to basic single- and multi-photon phenomena, such as Hong-Ou-Mandel interference and violation of the Bell-CHSH inequality, and confirm that the calculated properties are quantitatively comparable to the experimental results. Furthermore, our method visualizes the spatial propagation of photons hence provides insights that aid experiment designs for quantum-enabled technologies.
\end{abstract}


\maketitle

\section{Introduction}\label{sec1}
Quantum electrodynamics (QED) is a subset of quantum field theory~\cite{Weinberg} that describes the interaction between matter and the electromagnetic field. This theory treats light as particles called photons.
Arguably, QED is the most relevant theory considering the length, time, and energy scale required to observe the fundamental forces.
If the matter appears in the form of atoms and molecules, then the combined matter-light system becomes a quantum optical system \cite{Mandel}.
The dynamics of quantum optical systems, which we call quantum optical dynamics (QOD), is also governed by the basic equations of QED.
One of the most critical difficulties in solving the equations of QOD is the curse of dimensionality, which comes from the dimension (the size of a basis) of the Hilbert space corresponding to a quantum optical system. A basis can be characterized by ``modes", which roughly corresponds to the different ways in which particles (both matter and photons) can be excited, independently. The number of possible modes is infinite. Even if the number of modes is restricted, the size of the basis increases exponentially in the number of particles present in QOD.
A common simplification is to limit the number of modes and/or the total number of particles of each mode to just a few.
This approximation returns reasonable results if most contributing degrees of freedom can be confidently identified from previous experiments.
Should that fail, we have no other means but to numerically solve high-dimensional equations to accurately analyze QOD.

Known QOD simulation techniques \cite{Bello2019, phet, Langford2017, Lamata2020, qutip} fall into two categories:
cavity QED \cite{cavityQED} and wave packet dynamics \cite{Nysteen2015}.
In a typical cavity QED simulation, the photonic modes are restricted to a single standing wave mode in the cavity, and the matter-light interaction modeled by the Jaynes-Cummings Hamiltonian \cite{JCmodel1, JCmodel2}. This suffices to explain several physical phenomena such as vacuum Rabi oscillation.
On the other hand, the wave-packet-dynamics approach treats photons as a localized wave packet.
Previous works treat Gaussian-shaped photons propagating through one-dimensional waveguides and interacting with two-level atoms \cite{Nysteen2015, Hu2021, Stolyarov2019, Stolyarov2014, Stolyarov2013}. However, the restriction to the one-dimensional waveguide hinders their application to spatially two or three-dimensional systems.
Havukainen {\it et al.} \cite{Havukainen1999} conducted numerical simulations of wave packet dynamics where a single photon propagates through an ensemble of two-level atoms laid out in a two-dimensional space. 
The simulation divides the space into smaller grids and retains all the spatial modes up to the resolution determined by the size of the grids.
The two-level atoms are located at the grid points (see the top left panel of Fig.~\ref{fig:mz}(a)).
The dimension of the Hilbert space handled by this simulation exceeds that of the others by several orders of magnitude.
The high dimensionality allows this simulation to reproduce a variety of QOD such as reflection and interference of a single photon by mirrors, beamsplitters, and double slits.

All these previous methods address the curse of dimensionality by either limiting the spatial modes and/or the total number of particles.
For example, Havukainen {\it et al.} \cite{Havukainen1999} 
considered a large number of possible spatial modes, but they treated the dynamics of one photon.
The multi-photon system gives rise to characteristic quantum phenomena that involve quantum entanglement.
These phenomena are not only important for basic science, but also are the underpinnings of quantum-enabled applications in cryptography, sensing, and imaging~\cite{q_imaging1, q_imaging2, DIQKD_exp1, DIQKD_exp2}. Polarization is a degree of freedom of photons that is used in many quantum applications due to the availability of means to manipulate this degree of freedom at quantum precision. It is desirable that QOD simulations incorporate polarization.

In this paper, we introduce a numerical method to analyze multi-photon, spatially multi-dimensional QOD. Our method includes the polarization degree of freedom, based on the Hamiltonian in the quantum optical systems proposed by Havukainen {\it et al.} \cite{Havukainen1999}.
We construct the time evolution operator of the multi-photon system by composing a group of time evolution operators, each of which acting on a particular photon.
This treatment  exponentially reduces the dimension of the time evolution operator that needs to be computed, with respect to the number of photons (see Methods section for the details).
The numerical stability of simulations is improved by implementing a symplectic integration of the QED equations based on the Suzuki-Trotter decomposition. All combined, we succeed to simulate basic one- and two-photon phenomena, namely, the Mach-Zehnder (MZ) interference, Hong-Ou-Mandel (HOM) interference and violation of Bell-CHSH inequality.
We use this simulation method to visualize the photon propagation dynamics of these phenomena and to understand the physical origins of the computed results.
We also simulate a photon directed toward a scattering object (scatterer) and present the detailed interplay between the detection pattern and the single-photon interference caused by the scatterer.
The simulation results are presented in Sec. \ref{sec2}, followed by discussions in Sec. \ref{sec12}. 
The details of the present methods and parameters used in the numerical simulations are summarized in the Methods section.

\section{Results}\label{sec2}

\subsection{Mach-Zehnder interference} \label{subsec:mz}
\begin{figure}
 \centering
 \includegraphics[width=0.9\linewidth,bb=5 11 1508 780]{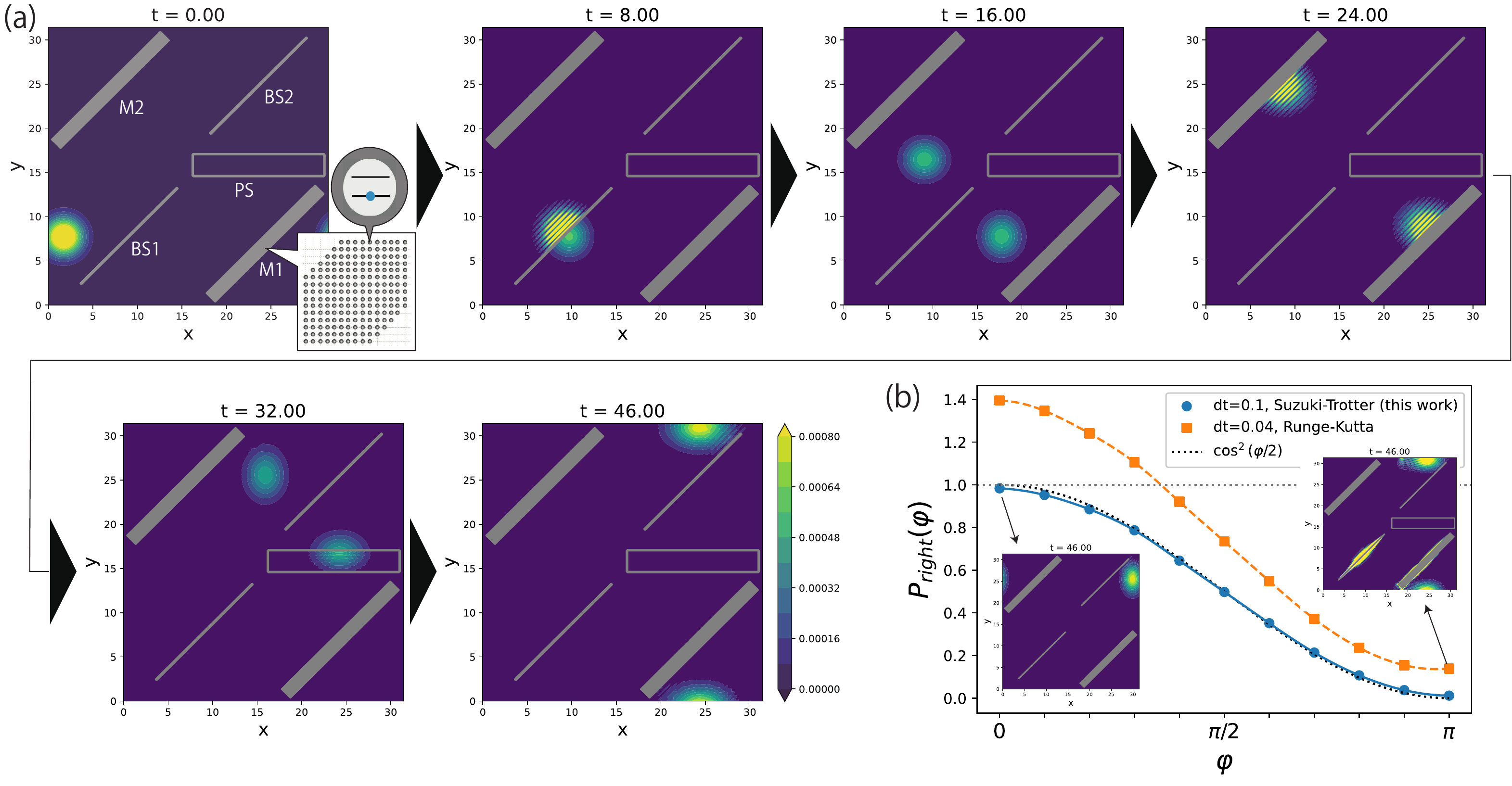}
  \caption{
  (a) Snapshots of the simulation of the MZ interference when imposed the phase shift $\varphi = \pi$.
  The contour shows the photon number density.
  The gray lines and filled rectangles correspond to the beamsplitters (BS1, BS2) and mirrors (M1, M2), respectively.
  The unfilled rectangle corresponds to the phase shifter (PS), which adds the phase-shift $\exp(i\varphi)$ to the photon. 
  (b) Plot of the probability $P_{\textrm{right}}(\varphi)$ obtained by the Suzuki-Trotter (blue) and the Runge-Kutta (orange) method. 
   $dt$ is the time step used in each method. 
  The black dotted curve shows the standard theoretical prediction. The horizontal gray dotted line is drawn at $P_{\textrm{right}} = 1$ as a guide for the eye.
  }
\label{fig:mz}
\end{figure}

The MZ interference is used for a variety of applications in optics, including optical switches \cite{opt_switch}, modulators \cite{opt_modulator}, sensors \cite{MZ_sensor}, and quantum computing \cite{MZ_computing}. 
A typical MZ interferometer uses a set of representative linear optical objects, namely, two beamsplitters (BS1, BS2), two mirrors (M1, M2) and one phase shifter (PS) deployed as shown in Fig.~\ref{fig:mz}(a).
The BS1 splits an incident beam of light into two beams, one which runs through M1 followed by PS and the other beam through M2. The two beams are then combined by BS2 to produce an interference pattern. 

We simulate the MZ interference of a single photon.
The optical objects are each implemented by an ensemble of two-level atoms, indicated by a gray filled rectangle in Fig.~\ref{fig:mz}(a).
The parameters of the two-level atoms are tuned so as to serve as the desired linear optical object (see Table \ref{tb:param_list} for specific parameters).
At $t=0$, a single photon is generated just left of BS1 directed toward BS1.
The probability $P_{\textrm{right}}$ that the photon is ejected from the right side of BS2 is determined by the phase shift $\varphi$ imposed by PS. A simplified theory predicts that $P_{\textrm{right}}(\varphi) = \cos^2(\varphi/2)$, assuming that a photon propagates in at most two modes at all times with no photons absorbed by the two-level atoms. 

Figure~\ref{fig:mz}(a) shows snapshots of the time evolution in the case of $\varphi = \pi$. 
Because the photon is ejected only in the upward direction, the result confirms the solution $P_{\textrm{right}}(\pi)=0$.
Figure~\ref{fig:mz}(b) shows a comparison of $P_{\textrm{right}}(\varphi)$ simulated by the Suzuki-Trotter decomposition and the Runge-Kutta method used in the present and previous works, respectively (see Methods section for the details).
The result of using the Suzuki-Trotter decomposition is in a good agreement with the theoretical prediction, despite the fact that more than two modes are involved in the process and photons are absorbed by the two-level atoms at the intermediate time steps.  On the other hand, the result of the Runge-Kutta method shows that the probabilities significantly deviate from the theoretical prediction, and some of the values exceed one. Furthermore, even using a shorter time step than that employed in the Suzuki-Trotter method, a numerical instability appears in QOD as shown in the inset of Fig.~\ref{fig:mz}(b). We conclude that the present method offers more reliable simulation than previously possible.

\subsection{Photon detection in the presence of scatterer} \label{subsec:scatterer}
\begin{figure}
 \centering
 \includegraphics[width=0.9\linewidth,bb=1 0 900 1007]{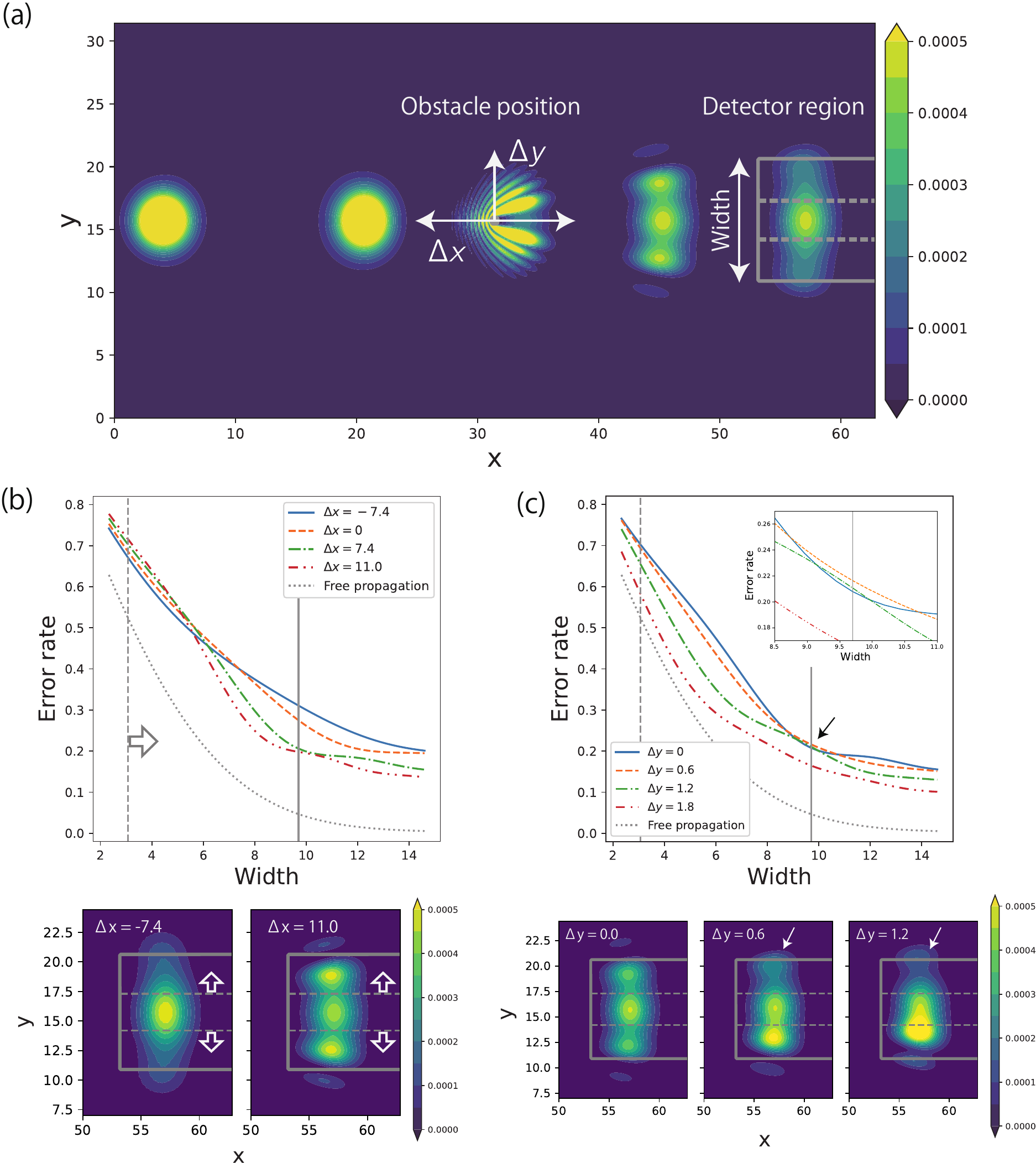}
  \caption{
  (a) The time evolution of the photon number density with the scattering obstacle located at the center of space. Two typical widths of the detector region are illustrated by the dashed and solid lines. The width between the two dashed lines is narrower and that between the two solid lines is wider.
  (b) The upper panel shows the error rate as a function of the width of the detector window. The results are obtained by changing $\Delta x$ but fixing $\Delta y = 0$. The gray dotted line shows the result without a scatterer.
  The vertical lines correspond to the typical widths of the detector region. 
  The lower panels show the final state of $\Delta x=-7.4$ and $11.0$.
  (c) The same as (b) but the plots denote the different values of $\Delta y$. We set $\Delta x = 7.4$. The inset shows detailed plots around the region indicated by the black arrow.
  The lower panels show the final states for $\Delta y=0.0, 0.6$, and $1.2$.
  }
\label{fig:err_rate}
\end{figure}

Given a practical use of a quantum sensing and imaging~\cite{q_imaging1, q_imaging2}, obstacles in a space may scatter photons and affect the detection accuracy.
Such a system cannot be described by an idealized theory, where the photonic modes are limited to one or just a few.
Here, we demonstrate scattering of a single photon by an obstacle (scatterer), as shown in Fig.~\ref{fig:err_rate}(a), exploiting the simulation capacity of our method for high-dimensional QOD. The scatterer, which is composed of $4 \times 4$ two-level atoms, is located at the center of the simulation space and is shifted by $\Delta x$ and $\Delta y$ in the $x$ and $y$ direction, respectively.
A detector is placed at the far right edge of the space, indicated by the gray rectangle.
The probability $P$ of the propagated photon being detected is given by the photon number density within the detector region.
The error rate is calculated by 1- $P$.
We computed the error rate for various values of the widths of the detector region.
In general, we expect that the scatterer should have less effect on the photon as the scatterer moves further away from the optical path. We shall see that this general intuition does not hold.

Figure~\ref{fig:err_rate}(b) shows the dependence of the error rates on the widths of the detector window.
We change $\Delta x$ but fix $\Delta y = 0.0$.
As a general trend, the error rates monotonously decrease as the width increases. This is because the wider the width of the detector region, the more likely it is to find the photon inside the detector region.
Yet, we observe a plateau profile of the error rate in $\Delta x=11.0$ around the width indicated by the vertical gray solid line. 
The number densities at the final states are visualized for $\Delta x=-7.4$ and $11.0$ in the lower panels of Fig.~\ref{fig:err_rate}(b).
When $\Delta x=11.0$, the number density has a particular structure, that is caused by an interference of a scattered photon coming from the upper and lower side of the scatterer. This fringe structure gradually disappears by diffusion of the photon number density as the distance to the scatterer increases after passing it, as in the case of $\Delta x=-7.4$. The plateau profile appears when an edge of the detector region is in the sparse region of the interference fringes. In a sense, the detector ``fails" to capture more photons despite increasing its size.

Figure~\ref{fig:err_rate}(c) compares the profiles of the error rates by changing $\Delta y$ values and fixing $\Delta x=7.4$. 
As in Fig. \ref{fig:err_rate}(b), the error rates generally decrease as the width increases.
When $\Delta y$ increases, the error rates decrease because the scatterer moves further away from the optical path.
We observe that the order of the error rate curves changes as the detector region width changes (cf. the black arrow).
The result shows that there is a range of the widths where the error rate increases when the scatterer moves away from the center of the optical path (also see the inset of Fig.~\ref{fig:err_rate}(c)).
To understand this
behavior,
we visualize the final states of $\Delta y=0.0, 0.6$, and $1.2$ as shown in the lower panels of Fig.~\ref{fig:err_rate}(c).
In the case of $\Delta y=0$, the interference fringes almost completely land within the detector region. 
On the other hand, a shift $\Delta y$ destroys the fringe structure.
Due to the cancellation of the interference, a part of the wave-packet distributes outside the detector region (indicated by the white arrows in the lower panels of Fig.~\ref{fig:err_rate}(c)), resulting in the higher error rate than the $\Delta y=0$ case.

\subsection{Hong-Ou-Mandel interference}\label{subsec:hom}
\begin{figure}
 \centering
 \includegraphics[width=0.9\linewidth,bb=1 4 473 453]{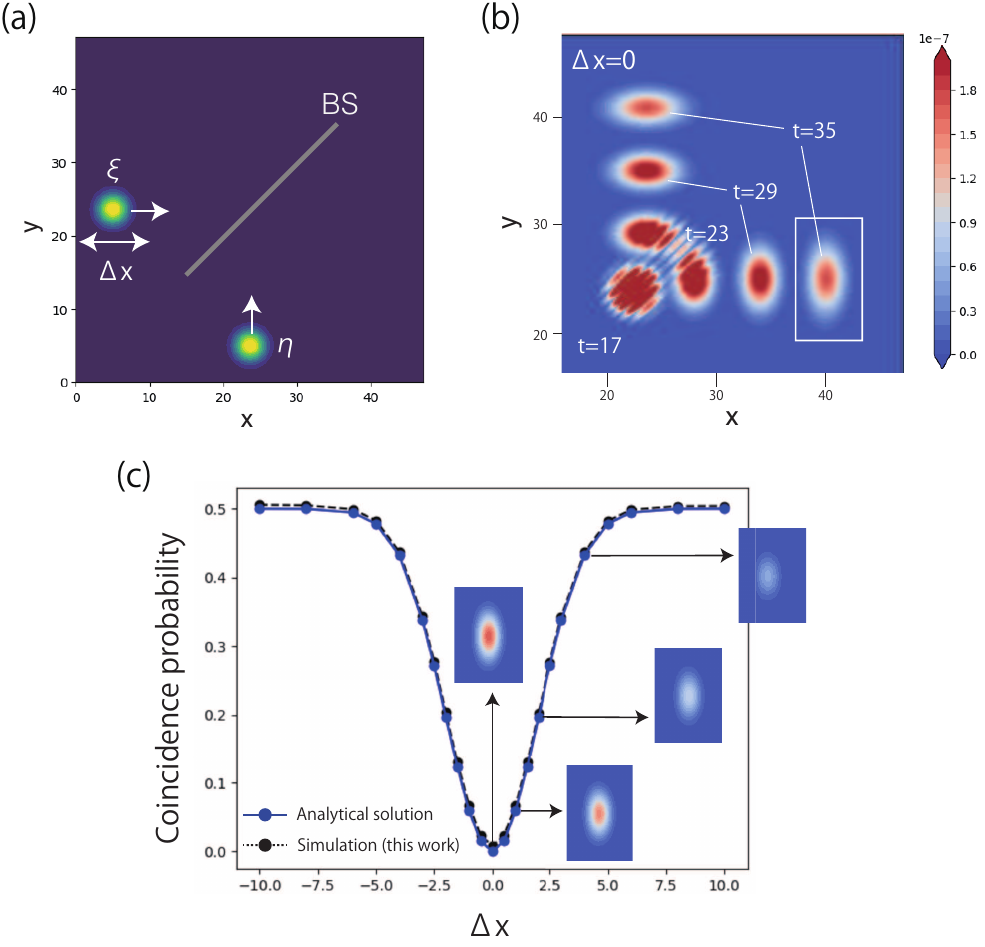}
  \caption{ (a) Schematic of a setup for HOM interference.
The beamsplitter is located at the center of the space tilted at 45 degrees.
The initial position of the $\xi$ photon in the $x$ direction is shifted by $\Delta x$.
(b) Time evolution of the probability distribution of the two photons being found at the same position, i.e., bunching probability distribution.
(c) Simulation result and analytical solution in Eq. \ref{eq:overlapinteg} of the HOM interference. 
The vertical axis indicates the coincidence probability, namely, the probability of detecting a photon simultaneously in the two separate output directions (up and right).
The inset panels show the bunching probability distribution within the white rectangle in (b).
  }
\label{fig:HOM}
\end{figure}

The Hong-Ou-Mandel (HOM) interference is a quintessential quantum effect of two indistinguishable photons \cite{HOM_original, HOM_review1}, which cannot be analyzed by simulations based on classical electrodynamics or QOD simulation limited to a single-photon states.
Figure \ref{fig:HOM}(a) shows a minimal model for the HOM interference, where two photons characterized by $\xi$ and $\eta$ indices are simultaneously injected to the ports of a beamsplitter. The beamsplitter divides a injected photon in half into transmitted and reflected parts. 
In QED, the injected state is expressed by
\begin{eqnarray}
 \ket{1_{\xi}, 1_{\eta}} = \hat{a}^{\dagger}_{\xi} \hat{a}^{\dagger}_{\eta} \ket{{0}},
\end{eqnarray}
where $\hat{a}^{\dagger}$ is the creation operator of a photon.
The operator of the beamsplitter $V_{\textrm{BS}}$ performs as,
\begin{eqnarray}
V_{\textrm{BS}} \hat{a}_\xi^\dagger V_{\textrm{BS}}^\dagger = \frac{1}{\sqrt{2}}(\hat{a}_\xi^\dagger + \img \hat{a}_\eta^\dagger ) \nm \\
V_{\textrm{BS}} \hat{a}_\eta^\dagger V_{\textrm{BS}}^\dagger = \frac{1}{\sqrt{2}}(\hat{a}_\eta^\dagger + \img \hat{a}_\xi^\dagger).\nm 
\end{eqnarray}
Therefore, the outgoing state from the beamsplitter becomes
\begin{eqnarray}
V_{\textrm{BS}} \otimes V_{\textrm{BS}}  \ket{1_{\xi}, 1_{\eta}}
= \frac{\img}{\sqrt{2}}(\ket{2_\xi, 0_\eta} + \ket{0_\xi, 2_\eta}) \label{eq:HOM_theory}.
\end{eqnarray}
This result indicates that the two photons are always ejected together from either right ($\xi$) or upper ($\eta$) outlet port.
In other words, the probability that each photon is emitted in a separate port disappears by quantum interference of the two photons.

This HOM interference is simulated by the model shown in Fig. \ref{fig:HOM}(a).
The calculated quantum state is denoted by $\ket{\Phi(\Delta x, t)}$, 
where $\Delta x$ is a shift in the x direction of the initial position of the $\xi$ photon.
First, we visualize the spatial distribution of the probability of finding two photons at a given location at different times of the evolution.
This bunching probability can be calculated by 
$\rho(\Delta x, t, \br) = |\bra{1_\br}\braket{1_\br | \Phi(\Delta x, t)}|^2$, 
where $\ket{1_\br}\ket{1_\br}$ is a state in which two photons are in the same position.
Figure \ref{fig:HOM}(b) shows the dynamics of $\rho(0, t, \br)$.
Before the two photons reach at the beamsplitter ($t < 17$), 
 there is almost no value of the bunching probability because they are far from each other.
After the photons pass through the beamsplitter, we observe that the bunching appears and then the distribution is separated into two portions that travel in the right and upper direction.
Then, we perform this numerical experiment for different values of $\Delta x$. 
We remove the beamsplitter in Fig. \ref{fig:HOM}(a) and calculate the two-photons system.
This free propagating state is denoted by  $\ket{\Phi_0 (t)}$ in which
the two photons are observed separately in the different outlets, which we call the coincidence probability.
The coincidence probability can be measured by
\begin{eqnarray}
p(\Delta x) = |\braket{\Phi_0 (T)|\Phi(\Delta x, T)}|^2,
\end{eqnarray}
where time $T = 45$ is chosen to assure that the wave-packets have separated enough after passing through the beamsplitter.
Figure \ref{fig:HOM}(c) shows the profile of $p(\Delta x)$. 
Around $\Delta x = 0$, the coincidence probability is almost zero which corresponds to the theoretical result in Eq. \ref{eq:HOM_theory}.
Increasing $\Delta x$, we observe a dip in the coincidence probability.
This HOM dip was observed in experiments as an evidence of the quantum nature of light \cite{HOM_original, HOM_review1}.
The shape of the dip is characterized by an overlap of the two photons. In fact, a theory of the HOM interference offers an analytical solution~\cite{HOM_original}.
\begin{eqnarray}
p(\Delta x) = \frac{1}{2} \Bigl( 1-\exp\Bigl(- \frac{\Delta x^2}{2\sigma^2} \Bigr)\Bigr), \label{eq:overlapinteg}
\end{eqnarray}
where $\sigma$ is the Gaussian width of the photon.
This analytical solution, which is plotted in Fig. \ref{fig:HOM}(c), shows a good agreement with the numerical one.

\subsection{Violation of Bell-CHSH inequality}\label{subsec:chsh}
In addition to the multi-photon states, the photon polarization degree of freedom is included in the present method. The polarization is commonly chosen in quantum experiments, including the photonic experiments confirming the violation of Bell-CHSH inequality \cite{Bell, CHSH}. The CHSH inequality provides a limit on a particular type of correlations of two separated systems, should their behavior be determined by classical mechanics. Quantum mechanics, however, violates the inequality under specific conditions. This violation was successfully confirmed by photonic experiments \cite{Clauser1972, Aspect1982, Zeilinger1998} and various physical systems including atoms \cite{Rosenfeld2017} and superconducting qubits \cite{Ansmann2009}.

\begin{figure}
 \centering
 \includegraphics[width=0.9\linewidth,bb=1 0 826 1018]{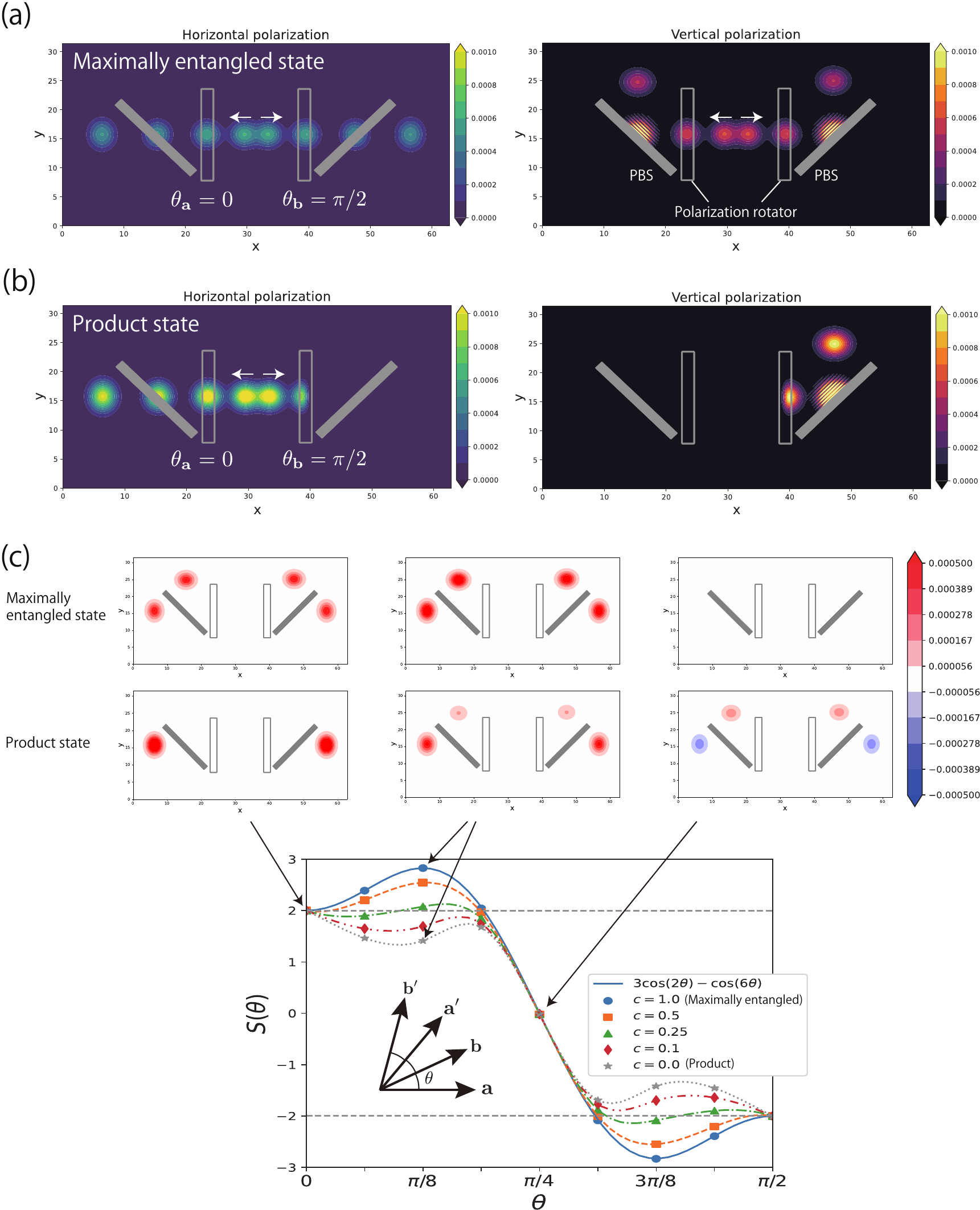}
  \caption{
  (a) Time evolution of two photons in the Bell-CHSH experiment, in the case of the maximally entangled state.
  The left and right panel show the photon number density of the horizontal and the vertical polarization component, respectively.
  The filled and unfilled rectangles represent the polarization beamsplitter (PBS) and the polarization rotator, respectively. The angles of the two polarization rotators are $\theta_{\mathbf{a}} = 0$, $\theta_{\mathbf{b}} = \pi/2$. 
  (b) The same as (a) but the input photon is a product state.
  (c) The correlation value $S(\theta)$ in Eq. \ref{eq:stheta}. 
  The horizontal dashed gray lines indicate the limit of the local realism. 
  The inset explains a configuration of the angles set in this simulation: $\theta_{\mathbf{a}}=0$ and $\theta_{\mathbf{ab}} = \theta_{\mathbf{a'b}} = \theta_{\mathbf{a'b'}} = \theta$.
  The upper panels show the correlation densities $S(\theta, \br)$ of the maximally entangled and product states at $\theta=0, \pi/8, \pi/4$.
  }
\label{fig:chsh}
\end{figure}

Here, we model the experiment performed by Aspect {\it et al.} \cite{Aspect1982_2}, as shown in Figs.~\ref{fig:chsh}(a) and (b).
We deploy optical objects that rotate the polarization of the photon passing through them.
One of the polarization rotators shifts by $\theta_{\mathbf{a}}$ or $\theta_{\mathbf{a}'}$, while the other by $\theta_{\mathbf{b}}$ or $\theta_{\mathbf{b}'}$.
Two photons are directed to the two rotators individually, and then the horizontal and vertical components are spatially separated by the polarization beamsplitters.
Given the rotation angles, we observe a correlation of the polarizations of the two photons as defined by
\begin{eqnarray}
S(\theta) = E(\theta_{
\mathbf{a}},\theta_{\mathbf{b}}) + E(\theta_{\mathbf{a}'},\theta_{\mathbf{b}}) + E(\theta_{\mathbf{a}'},\theta_{\mathbf{b}'}) - E(\theta_{\mathbf{a}},\theta_{\mathbf{b}'}), \label{eq:stheta}
\end{eqnarray}
where $E(\theta_{\mathbf{a}},\theta_{\mathbf{b}}) = P_{HH}(\theta_{\mathbf{a}}, \theta_{\mathbf{b}}) + P_{VV}(\theta_{\mathbf{a}}, \theta_{\mathbf{b}}) - P_{HV}(\theta_{\mathbf{a}}, \theta_{\mathbf{b}}) - P_{VH}(\theta_{\mathbf{a}}, \theta_{\mathbf{b}})$.
The probability $P_{pp'}(\theta_{\mathbf{a}}, \theta_{\mathbf{b}})$ ($p,p' \in \{H,V\}$) is the probability that a photon is detected in the $p$-polarization outlet in the left and, simultaneously, another photon detected in the $p'$-polarization outlet, when the rotation angles are set at $\theta_{\mathbf{a}}$ and $\theta_{\mathbf{b}}$.
According to the Bell-CHSH inequality, $S(\theta)$ must range $-2 \leq S(\theta) \leq 2$ if the local realism is correct.
However, if we prepare two photons that are in a maximally entangled state, quantum mechanics asserts that
\begin{align}
P_{pp'} (\theta_{\mathbf{a}}, \theta_{\mathbf{b}}) = \frac{1}{2} |\delta_{p,p'}- \cos^2\theta_{\mathbf{ab}}|, \label{eq:P_pp}
\end{align}
where $\theta_{\mathbf{ab}}= |\theta_{\mathbf{a}} - \theta_{\mathbf{b}}|$ and 
$\delta$ is the Kronecker delta \cite{CHSH_eq}.
Equation \ref{eq:P_pp} leads to  $E(\theta_{\mathbf{a}},\theta_{\mathbf{b}}) = \cos 2\theta_{\mathbf{ab}}$.
When setting the angles at $\theta_{\mathbf{ab}} = \theta_{\mathbf{a'b}} = \theta_{\mathbf{a'b'}} = \theta$ (see also the inset of Fig.~\ref{fig:chsh}(c) ),
and hence $\theta_{\mathbf{ab'}} = 3\theta$, Eq. \ref{eq:stheta} becomes
\begin{align}
S(\theta) = 3\cos 2\theta - \cos 6\theta, \label{eq:theory_s}
\end{align}
in the case of the maximally entangled state.
Equation \ref{eq:theory_s} breaks the inequality $-2 \leq S(\theta) \leq 2$ as shown by the blue solid line of  Fig.~\ref{fig:chsh}(c).

Figure~\ref{fig:chsh}(a) shows the dynamics of the number density of two photons prepared in a maximally entangled state. Each photon propagates to the left and right direction, as indicated by the white arrows.
The angles of the two polarization rotators are set at $\theta_{\bf a}=0$ and $\theta_{\bf b}=\pi/2$. 
The initial state is given by $\ket{\Phi(t=0)}=(\ket{H}\ket{H}+\ket{V}\ket{V})/\sqrt{2}$, where $\ket{H}\ket{H}$ ($\ket{V}\ket{V}$) denotes the two photons polarized in the horizontal (vertical) direction.   
Figure~\ref{fig:chsh}(b) shows the case of a product state  $\ket{\Phi(t=0)}=\ket{H}\ket{H}$.  The photons of the product change their polarization angle by passing through the polarization rotator, while not for those of the maximally entangled state. This invariance can be explained by computing the state of one of the photons when the initial state of the two photons is $\ket{\Phi} = \frac{\ket{H}\ket{H} + c\ket{V}\ket{V}}{\sqrt{1+c^2}}$, where $0 \le c \le 1$.
The density matrix corresponding to the two photons is
\begin{equation}
\hat{\rho}(c) = \ket{\Phi}\bra{\Phi}. \label{eq:cshs_phi}
\end{equation}
A partial trace of the density matrix on the right photon yields the reduced density matrix
\begin{eqnarray}
\hat{\rho}'(c) = \sum_p \bra{p} \hat{\rho}(c) \ket{p}= \frac{1}{1+c^2} \Bigl( \ket{H}\bra{H} + c^2 \ket{V}\bra{V} \Bigr). \nm 
\end{eqnarray}
If $c=1$ which corresponds to the maximally entangled state, $\hat{\rho}'(1) = \hat{I}/2$ where 
$\hat{I}$ is the identity operator.
Considering a polarization rotation $\hat{U}$ that operates $\hat{U} \rho'(c) \hat{U}^{\dagger}$, it is obvious that the polarization rotator does not change the partial state of this photon.
On the other hand, we have $\hat{\rho}'(0) = \ket{H}\bra{H}$, thus the state of the photon of the product state is changed by the operation of $\hat{U}$.
These results are visualized in Figs.~\ref{fig:chsh}(a) and (b).

Figure~\ref{fig:chsh}(c) shows the simulated $S(\theta)$.
The results of the maximally entangled state (the blue plots) show a good agreement with the theoretical prediction (the blue solid line) of Eq. \ref{eq:theory_s}.
The violations of the Bell-CHSH inequality appear around $\theta=\pi/8,\ 3\pi/8$. In addition, we calculate $S(\theta)$ by changing the parameter $c$ of the initial state as in Eq. \ref{eq:cshs_phi}. The violations occur if  $c \ge 0.25$.

To obtain deeper insights into the difference between the maximally entangled and product states, we visualize $S(\theta)$ in Fig \ref{fig:chsh}(c). 
Namely, we first define 
\begin{align}
P_{pp'}(\theta_{\mathbf{a}}, \theta_{\mathbf{b}}, \br)
= \frac{1}{2} \sum_{\br'}
| \left(\bra{1_{\br,p}}\bra{1_{\br',p'}} + \bra{1_{\br',p}}\bra{1_{\br,p'}}  \right)  \ket{\Phi(t)}|^2 , \nm
\end{align}
and  $E(\theta_{\mathbf{a}},\theta_{\mathbf{b}}, \br) = P_{HH}(\theta_{\mathbf{a}}, \theta_{\mathbf{b}}, \br) + P_{VV}(\theta_{\mathbf{a}}, \theta_{\mathbf{b}}, \br) - P_{HV}(\theta_{\mathbf{a}}, \theta_{\mathbf{b}}, \br) - P_{VH}(\theta_{\mathbf{a}}, \theta_{\mathbf{b}}, \br)$. 
The basis $\ket{1_{\br,p}}$ is a state in which a photon with the polarization $p$ is at the position \br.
Accordingly, a correlation density is defined by
\begin{align}
S(\theta, \br) = E(\theta_{\mathbf{a}},\theta_{\mathbf{b}}, \br) + E(\theta_{\mathbf{a}'},\theta_{\mathbf{b}}, \br) + E(\theta_{\mathbf{a}'},\theta_{\mathbf{b}'}, \br) - E(\theta_{\mathbf{a}},\theta_{\mathbf{b}'}, \br). 
\end{align}
Note that we can retrieve $S(\theta)$ by $S(\theta) = \sum_\br S(\theta,\br)$.
The distribution of $S( \theta,\br)$ for the maximally entangled state is found to be invariant with respect to $\theta$ except for its magnitude.
This result is relevant to Fig.~\ref{fig:chsh}(a) in which the number density apparently does not change by the polarization rotators due to the fact that the reduced density matrix becomes the identity operator. Therefore, the correlations between the sets of $\theta_
{\mathbf{a}}$ and $\theta_{\mathbf{b}}$ tend to be magnified, resulting in clear violations of the inequality at particular angles. On the other hand, since the product state is changed by the polarization rotator as shown in Fig.~\ref{fig:chsh}(b), terms composed of the correlations tend to cancel each other. This fact prevents $S(\theta)$ to exceed the limit of the local realism.
In short, the present method reproduces the theory and experimental results, and its visualization facilitates understanding of the mechanism.

\section{Discussion}\label{sec12}
The curse of dimensionality does not exist if the system is modeled and solved by classical electrodynamics. Many aspects of optics can be analyzed by classical methods of electromagnetic fields, such as method of moments~\cite{MoM}, finite-difference time-domain methods~\cite{FDTD1, FDTD2, FDTD3}, finite-difference frequency-domain methods~\cite{FDFD}, and finite element method~\cite{FEM}.  These techniques are based on the Maxwell equations that approximate light-matter interactions as a dissipation of the electromagnetic field or as spatially and spectrally nonuniform permittivity and/or permeability (so-called macroscopic treatment).
However, the classical framework cannot capture multi-photon phenomena (see the recent review~\cite{Barnett2022}), such as quantum interference and quantum entanglement, which are demonstrated in this study. 
By modeling a three-level atomic system, 
our numerical scheme can simulate nonlinear optical effects such as spontaneous parametric down conversion. However, the curse of dimensionality would reappear because a direct product form of the time evolution operator is no longer applicable. We will tackle this in a future work.

While any QOD simulation has to contend with some omission of details when used to understand experiments,
visualization of quantum dynamics of complex systems, as treated in this paper, is a unique advantage of numerical simulation, not present in experiments. A quantum state changes itself in principle when observed. It is, therefore, significantly resource intensive to experimentally trace the time-evolution process of photons propagating through complex optical setups. The presented high-dimensional QOD would be useful to reveal the intermediate process in detail. For instance, the visualizations of Sec. \ref{subsec:scatterer} helped us to clarify the cause of the behavior of the error rate. This flexible and accurate simulation of the high-dimensional QOD should also aid in creating a viable minimal model that guides designing experiments and applications that exploit quantum states of light.

\clearpage

\section{Methods}\label{sec11}
We explain the present method in a two-dimensional space that spans over
$ 0 \le x \le L_x$ and $ 0 \le y \le L_y$.
The space is discretized by numbers of grids ($M_x$, $M_y$), and its boundaries are periodic.
We set $\hbar = c = 1$ in the following formulations for a notational simplicity.

\subsection{Single-photon system}
We begin with a theoretical framework of the single-photon simulation as introduced by \cite{Havukainen1999} (also see the Supplemental Information for the details).
The system consists of 
 $N_A$ two-level atoms (we write ``atom" in the following for the simplicity) and one photon that has a wave-number vector
$\bk$.
The total Hamiltonian is
\begin{eqnarray}
\hat{h} &=& \hat{h}_0 + \hat{h}_I \nm, 
\end{eqnarray}
where
\begin{eqnarray}
\hat{h}_0 &=&  \sum_{\bk} \omega_{\bk} \hat{a}_{\bk}^{\dagger} \hat{a}_{\bk} 
              + \sum_{j=1}^{N_A}  2 \omega_j \hat{a}_{j}^{\dagger} \hat{a}_{j}, \label{eq:h_0}
\end{eqnarray} 
and $\hat{a}$ is an annihilation operator.
The frequencies $\omega_{\bk}$ and $\omega_{j}$ indicate eigenenergies of the corresponding photon mode and atom, respectively.
The Hamiltonian $\hat{h}_I$ represents a dipole-dipole interaction between the photon and atoms,
\begin{eqnarray}
\hat{h}_I &=& \sum_{j, \bk} 
(
    g(j, \bk) \hat{a}_{j}^{\dagger}  \hat{a}_{\bk} 
+ g^*(j, \bk) \hat{a}_{\bk}^{\dagger} \hat{a}_{j}
) 
\label{eq:h_I} \\
g(j, \bk)  &=& -\frac{\img}{\sqrt{2}L} \sqrt{\omega_j} D_j e^{\img \bk \cdot \br_j},
\label{eq:g_jk}
\end{eqnarray} 
where $L= \sqrt{L_x L_y}$.
The parameter $D_j$ sets the strength of the dipole-dipole interaction, 
and the vector $\br_j$ the position of the $j$th atom.
This interaction makes the constituent atoms play a role of linear optical objects such as mirror, beamsplitter, and scatterer by controlling the parameters $\omega_j$ and $D_j$.

A quantum state of this system is restricted so that the total number of excitations is one, namely,
\begin{eqnarray}
\ket{\phi(t)} &=&  
    \sum_{\bk} c(t,\bk) \ket{1_{\bk}} + \sum_{j=1}^{N_A} c_{j}(t) \ket{1_{j}}, \nm
\end{eqnarray} 
where $c(t, \bk)$ and $c_{j}(t)$ are the complex amplitudes satisfying $\sum_{\bk} |c(t, \bk)|^2 + \sum_{j} |c_{j}(t)|^2  = 1$.
The atoms are deployed at the grid lattice points to form necessary optical objects.
The indices $1_{\bk}$ and $1_j$ in the kets denote one photon of $\bk$ mode and one excitation  of a $j$-th atom, respectively.
Namely, operations of $\hat{a}_\bk$ to the bases are
\begin{eqnarray}
\hat{a}_\bk \ket{1_{\bk}}=\ket{0}, \ \ \ \hat{a}^{\dagger}_\bk \ket{0} = \ket{1_{\bk}}, \nm 
\end{eqnarray}
and those of $\hat{a}_j$ are the same manner.
In addition, 
\begin{eqnarray}
\hat{a}^{\dagger}_j \hat{a}^{\dagger}_\bk \ket{0}  = \hat{a}^{\dagger}_\bk \hat{a}^{\dagger}_j \ket{0} = 0, \nm
\end{eqnarray}
due to the restriction of the total number of excitations to one.

The Schr$\ddot{\textrm{o}}$dinger equation and the solution of the time evolution are written by
\begin{eqnarray}
\img \frac{\partial \ket{\phi(t)}}{\partial t} &=& \hat{h} \ket{\phi(t)}    \label{eq:schrodinger}\\
\ket{\phi(t)} &=& \exp(- \img \hat{h} t) \ket{\phi(0)}. \label{eq:time_evolution_operator}
\end{eqnarray}

\subsection{Polarization degrees of freedom} \label{subsec:Polarization}
The method is extended simply by adding a polarization index $p$ that specifies the horizontal and vertical polarization by $H$ and $V$, respectively.
The state of the single-photon system is expressed by
\begin{eqnarray}
\ket{\phi(t)} &=& \sum_{p \in \{H, V\}}\ket{\phi_p(t)}  \label{eq:pol_1photon_state} \\
\ket{\phi_p(t)} &=& \sum_{\bk} c_p(t,\bk) \ket{1_{\bk, p}} + \sum_{j=1}^{N_A} c_{j, p}(t) \ket{1_{j, p}}. \nm
\end{eqnarray} 
The Hamiltonians $\hat{h}_0$ and $\hat{h}_I$ in Eqs. \ref{eq:h_0} and \ref{eq:h_I} can be extended 
by just replacing $\hat{a}_{\bk} \rightarrow \hat{a}_{\bk,p}, \hat{a}_{j} \rightarrow \hat{a}_{j,p}$.
Note that these Hamiltonians do not influence the polarization degrees of freedom.
The interaction Hamiltonian $\hat{h}_I$ becomes
\begin{eqnarray}
\hat{h}_I &=& \sum_{p, j, \bk} 
(
    g_p(j, \bk) \hat{a}_{j,p}^{\dagger}  \hat{a}_{\bk ,p} 
+ g_p^*(j, \bk) \hat{a}_{\bk ,p}^{\dagger} \hat{a}_{j,p}
) 
\label{eq:h_I_pol} \\
g_p(j, \bk)  &=& -\frac{\img}{\sqrt{2}L} \sqrt{\omega_{j}} D_{j,p} e^{\img \bk \cdot \br_j},
\label{eq:g_jk_pol}
\end{eqnarray} 
where $D_{j,p}$ denotes the strength of the dipole interaction between the $j$-th atom  and the photon mode $p$.
Tuning $D_{j,p}$ creates optical objects that change the polarization.
For example, a polarization beamsplitter reflects only the vertical polarization but transmits the horizontal one. 
It can be created by setting $\omega_{j}$ and $D_{j,V}$ at appropriate values for a role of mirrors, but  $D_{j,H}=0$.
A polarization rotator can be constructed by a rotated half wave plate which introduces $\pi$ phase shift on the polarization directed toward the slow axis of the wave plate. This can be achieved by using the basis of the fast and slow axes of the wave plate $\{ F,S\}$, as
\begin{align}
\ket{1_{\bk, F}} &= \cos\Theta\ket{1_{\bk, H}} + \sin\Theta\ket{1_{\bk, V}}, \ 
\ket{1_{\bk, S}} = -\sin\Theta\ket{1_{\bk, H}} + \cos\Theta\ket{1_{\bk, V}}, \nm \\
\ket{1_{j, F}}   &= \cos\Theta\ket{1_{j, H}} + \sin\Theta\ket{1_{j, V}}, \ 
\ket{1_{j, S}}   = -\sin\Theta\ket{1_{j, H}} + \cos\Theta\ket{1_{j, V}}. \nm
\end{align}
The basis transformation of the single-photon system is
\begin{align}
\sum_{\bk, p' \in \{F,S\}} \ket{1_{\bk, p'}} \braket{1_{\bk, p'}|\phi(t)} + \sum_{j, p' \in \{F,S\}}  \ket{1_{j, p'}} \braket{1_{j, p'}|\phi(t)}.
\end{align}
Then, $\Theta$ is set as $\Theta = \theta_{\textrm{rot}}/2 + \theta_{\textrm{pol}}$ depending on a rotation angle $\theta_{\textrm{rot}}$ given by the polarization rotator, where $\theta_{\textrm{pol}}$ denotes an angle of the linear polarization of the input photon against a direction of the horizontal polarization.
$\omega_{j}$ and  $D_{j,S}$ are set at appropriate values to add $\pi$ phase shift on the slow axis component, while $D_{j,F}=0$.

\begin{figure*}
 \centering
 \includegraphics[width=0.8\linewidth,bb=1 8 927 485]{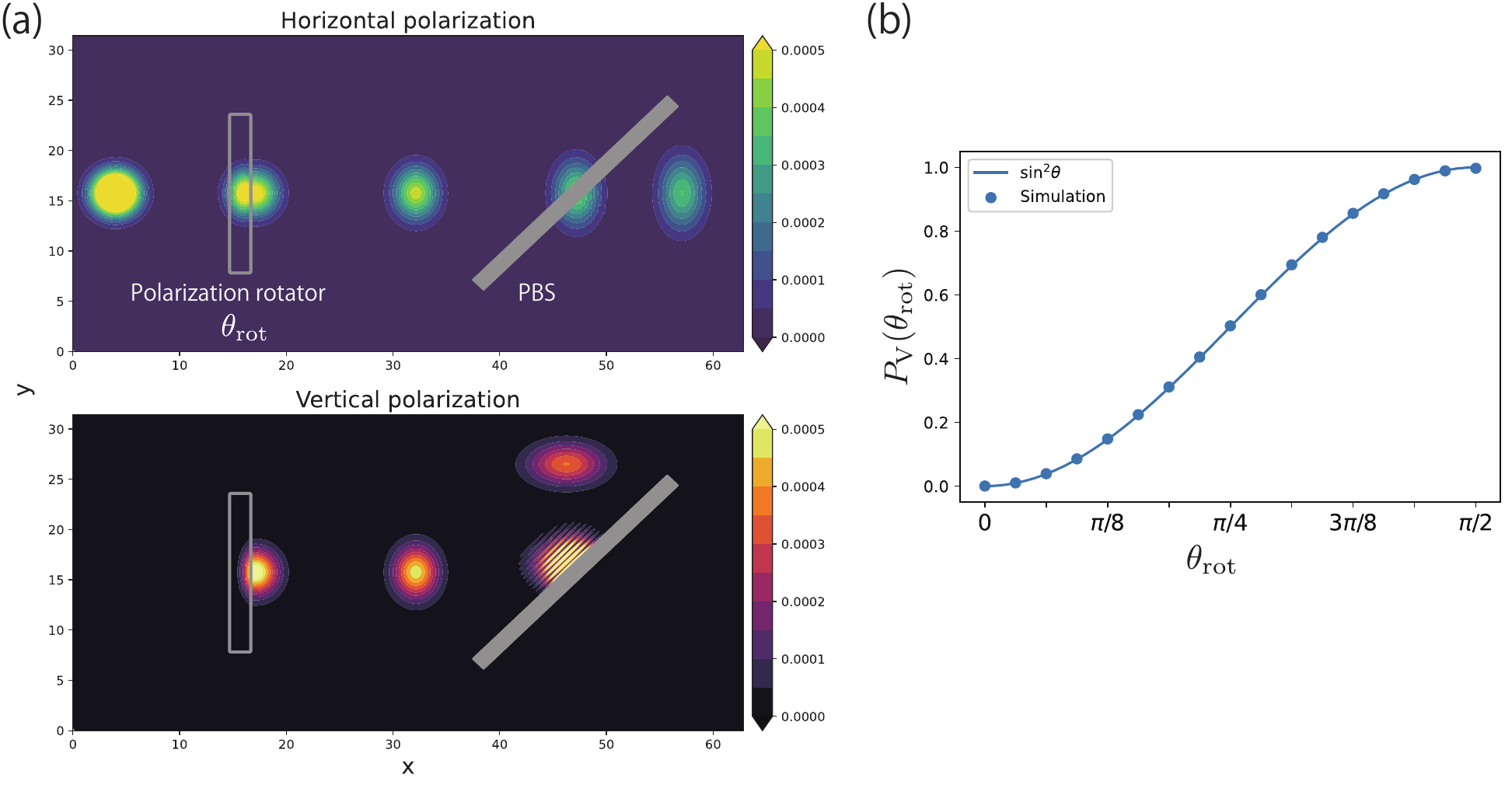}
  \caption{
  (a) The result of the simulation with the polarization rotator and the polarization beamsplitter (PBS). We set the rotation angle of the polarization rotator as $\theta_{\rm rot} = \pi/4$. The polarization rotator consists of 2048 atoms composed with 16 atomic layers, and its parameters are set as $D_{j,S}=0.56$ and $\omega_{j}=1.2$. PBS consists of 1185 atoms composed with 8 atomic layers, and its parameters are set as $D_{j,V}=0.56$ and $\omega_{j}=5.0$. The cavity lengths are $L_x=20\pi$ and $L_y=10\pi$, and the grids are cut by $512 \times 256$. The Gaussian width of the photon is $\sigma = 2.0$. The initial position and wave number of the photon are $\bar{\br} = (2.0, L_y/2)$ and $\bar{\bk}=(10.0, 0)$.
  (b) Plot of the probability $P_V(\theta_{\rm rot})$ which is the probability of the photon polarized in the vertical direction at the end of simulation. The solid line is a theoretical prediction assuming an idealized PBS.
  }
\label{fig:pol}
\end{figure*}

This scheme for the  polarization is examined by a single-photon simulation.
The time evolution is solved by the Suzuki-Trotter decomposition which will be explained in the Sec. \ref{subsec:suzuki-trotter}.
Figure \ref{fig:pol}(a) shows the system where the initial polarization of a photon is $H$, and the photon is directed toward a polarization rotator. In this case, the polarization is rotated by an angle $\theta_{\textrm{rot}}=\pi/4$. Then, only the $V$ component is reflected to the upward by the polarization beamsplitter, while the $H$ component passes through the object. 
We observe the probabilities of the $V$ components as a function of $\theta_{\textrm{rot}}$, as shown in Fig. \ref{fig:pol}(b).
The calculated probabilities are well-fitted by the behavior of an ideal polarization rotator $\sin^2 \theta_{\textrm{rot}}$.
These results confirm that the quantum state including the polarization and the relevant optical objects used in Sec.~\ref{subsec:chsh} work as intended.

\subsection{Time evolution by Suzuki-Trotter decomposition}\label{subsec:suzuki-trotter}
The previous study \cite{Havukainen1999} rewrites the Schr$\ddot{\textrm{o}}$dinger equation in Eq.~\ref{eq:schrodinger}
in the interaction picture and numerically solves it by the fourth-order Runge-Kutta method.
We have decided to formulate the Hamiltonian in such a way to apply the Suzuki-Trotter decomposition as it is known that the Suzuki-Trotter decomposition has numerically advantages in computing the time evolution of a Hamiltonian system \cite{MSuzuki_1990, MSuzuki_1993, DRSkeel_1997, YMiyamoto_2008, SGItoh_2013, MSuzuki_1976, NHatano_2005}.
The time evolution operator in Eq. \ref{eq:time_evolution_operator} can be approximated by the Suzuki-Trotter decomposition, as
\begin{eqnarray}
e^{-\img \hat{h}\delta t }  &=& e^{-\img (\hat{h}_0 + \hat{h}_I)\delta t } \nm \\
&\sim& e^{-\img \hat{h}_I \delta t /2} e^{-\img \hat{h}_0 \delta t } e^{-\img \hat{h}_I \delta t /2}, \label{eq:suzuki-trotter}
\end{eqnarray}
where $\delta t$ is a time step of the simulation.
Each Hamiltonian is diagonalized to compute the time evolution operator.
The $\hat{h}_0$ is already diagonal in the wave-number basis $\bk$.
To diagonalize $\hat{h}_I$, it is numerically convenient to transform the interaction Hamitonian $\hat{h}_I$ from the basis $\bk$ into the position basis $\br$
\begin{eqnarray}
\hat{h}_I = \sum_{p, j} (W_{j,p} \hat{a}_{j,p}^{\dagger}  \hat{a}_{\br_j, p} 
+ W^{*}_{j,p} \hat{a}_{j,p}  \hat{a}^{\dagger}_{\br_j, p}  ) , \label{eq:h_I_r}
\end{eqnarray}
because by using a Fourier transformation the term in Eq.~\ref{eq:h_I} becomes
\begin{eqnarray}
\sum_{\bk} g_p(j, \bk) \hat{a}^{\dagger}_{j,p}  \hat{a}_{\bk, p} 
&=& -\frac{\img}{\sqrt{2}L} \sqrt{\omega_j} D_{j,p} \hat{a}_{j,p}^{\dagger}  \sum_{\bk}  e^{\img \bk \cdot \br_j} \hat{a}_{\bk, p} \nm \\
&=&  W_{j,p}  \hat{a}_{j, p}^{\dagger}  \hat{a}_{\br_j, p}, \nm
\end{eqnarray}
where $W_{j,p} = -\img D_{j,p}  \sqrt{M \omega_{j}}/\sqrt{2} L$.
$M = M_x M_y$ is the number of grids.
The operation of $\hat{a}_{\br, p}$ is
\begin{eqnarray}
\hat{a}_{\br, p} \ket{1_{\br, p}}=\ket{0}, \ \ \ \hat{a}^{\dagger}_{\br,p} \ket{0} = \ket{1_{\br, p}}. \nm 
\end{eqnarray}
As seen in Eq. \ref{eq:h_I_r}, $\hat{h}_I$
exchanges  the photon and the excitation of the atom located at the same position.
This compact representation thanks to the $\br$ basis is advantageous in the numerical diagonalization.
In fact, multiplying $\ket{\phi}$ in Eq. \ref{eq:pol_1photon_state} gives
\begin{eqnarray}
\bra{1_{j,p}}   \hat{h}_I  \ket{\phi(t)} &=& W_{j,p} c_p(t, \br_j) \nm \\
\bra{1_{\br_j,p}} \hat{h}_I  \ket{\phi(t)} &=& W_{j,p}^{*} c_{j,p}(t). \nm
\end{eqnarray}
By using matrix representation, this is equivalent to
\begin{eqnarray} 
\hat{h}_I \vec{\phi}_p(t) &=&
\begin{pmatrix}
0        & W_{j,p}^*  \\
W_{j,p} & 0 
\end{pmatrix}
\begin{pmatrix}
c_p(t,\br_j)  \\
 c_{j,p}(t) 
\end{pmatrix}
\nm \\
&=&
V^{-1}
\begin{pmatrix}
-\img W_{j,p} & 0   \\
0 & \img W_{j,p} 
\end{pmatrix}
V
\begin{pmatrix}
c_p(t,\br_j)  \\
 c_{j,p}(t)
\end{pmatrix}
\nm 
\end{eqnarray}
where
$ V = 2^{-1/2}
\begin{pmatrix}
1 & -\img  \\
1 & \img 
\end{pmatrix}
$
is a unitary matrix and we used the fact that $W_{j,p}^* = -W_{j,p}$. Therefore, the time evolution can be operated, as
\begin{eqnarray} 
e^{-\img \hat{h}_I \delta t} \vec{\phi}_p(t) &=&
V^{-1}
\begin{pmatrix}
e^{- W_{j,p} \delta t}        & 0   \\
0 & e^{ W_{j,p} \delta t}
\end{pmatrix}
V
\begin{pmatrix}
c_p(t,\br_j)  \\
 c_{j,p}(t)
\end{pmatrix}
.\nm 
\end{eqnarray}

\begin{figure*}
 \centering
 \includegraphics[width=1.0\linewidth,bb=0 2 480 139]{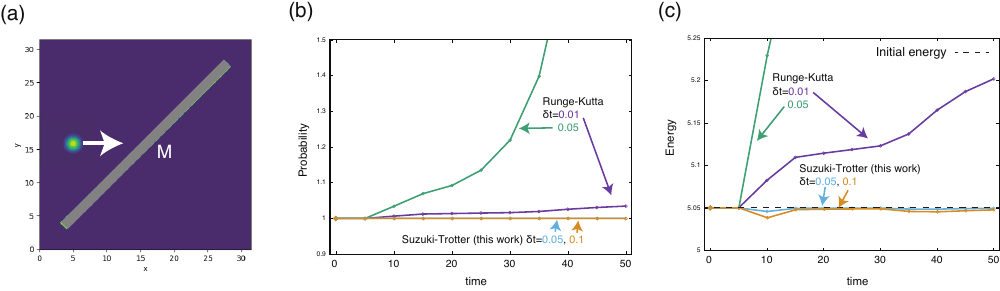}
  \caption{(a) Schematics of a test system.
  The space lengths are $L_x=L_y=10\pi$ and the grids are $256 \times 256$. The boundary is periodic.
The mirror consists of 1,584 atoms ($D_j=0.5, \omega_j=2.5$), and the width of mirror is composed of 8 atomic layers.
The mirror denoted by ``M" is located at the center of space with tilted at 45 degrees. The initial position and wave number of the injected photon are $\bar{\br} = (5.0, L_y/2)$ and $\bar{\bk}=(5.0, 0)$. The width is $\sigma = 1.0$.
Trajectories of (b) probability of the quantum state defined by $\braket{\phi(t)|\phi(t)}$ and (c) total energy $\bra{\phi(t)} \hat{h}\ket{\phi(t)}$. The present method using Suzuki-Trotter decomposition is compared with the method used in \cite{Havukainen1999}, namely, the fourth-order Runge-Kutta in the interaction picture.
  }
\label{fig:test_time_evolution}
\end{figure*}

We assessed a numerical performance of the Suzuki-Trotter decomposition by a simple single-photon simulation without the polarization degree of freedom.
As shown in Fig. \ref{fig:test_time_evolution}(a),
the injected photon collides with the tilted mirror at about $t=10$ and is reflected upwards. Then, due to the periodic boundary condition, the photon emerges from the lower region and is reflected by the mirror again at about $t=35$ to go in the right direction.
Figure \ref{fig:test_time_evolution}(b) shows the trajectories of the probability $\braket{\phi(t)|\phi(t)}$ that ideally should remain $1$ all the time.
The result obtained by the previous scheme based on the Runge-Kutta method increases significantly above $1$.
On the other hand, the present scheme keeps the probability at $1$, even with a larger time step.
This is a benefit of the Suzuki-Trotter decomposition in Eq. \ref{eq:suzuki-trotter} that uses only unitary operators.
Figure \ref{fig:test_time_evolution}(c) shows the total energy of the quantum state.
By using the previous scheme, the total energy increases and violates the law of energy conservation.
The energy trajectory by the present scheme shows small deviations from the initial energy at the times when the photon interacts with the mirror. 
Afterwards, however, the energy recovers to the initial energy, and the energy conservation tends to be maintained in the dynamics.
Such a favourable feature is known as the result of a symplectic condition
the Suzuki-Trotter decomposition has \cite{YMiyamoto_2008, SGItoh_2013}.
The above observations suggest that the time step $\delta t = 0.1$ is sufficiently small to justify the approximation of Eq.~\ref{eq:suzuki-trotter}.
Although an error of a time step in the second-order Suzuki-Trotter decomposition is an order of $\delta t^3$ which is larger than that of the fourth-order Runge-Kutta method ($\delta t^5$), its property of the energy conservation hinders the accumulation of the error in the long-time dynamics \cite{Garanin2021}, making the method preferable. To keep the error of the Runge-Kutta method comparable to that of the Suzuki-Trotter method, a significantly smaller time step than that used in the Suzuki-Trotter method is required (see Fig. \ref{fig:test_time_evolution}(c)).
Such a small time step leads to the longer computational time for the Runge-Kutta method.
The time evolution by the Suzuki-Trotter decomposition significantly improves stability of the calculation, which gives a reliable quantitative evaluation by the QOD simulation.

\subsection{Multi-photon system}
For notational simplicity, we omit the polarization index $p$, since the inclusion of $p$ can be done by replacing $\br \rightarrow \br, p$ and $j \rightarrow j, p$.
When naively written, an $N$-photon state without the polarization is
\begin{equation*}
	\ket{\Phi(t)} = \sum_{\br_1, \cdots \br_N} c(t, \br_1, \cdots \br_N) 
 \ket{1_{\br_1}}  \cdots \ket{1_{\br_N}} + \sum_{j} c_{j}(t) \ket{1_{j}}.
\end{equation*} 
A straightforward updating of the coefficients $c(t, \br_1, \br_2, \cdots \br_N)$ would require to represent the time evolution operator as an $O(M^N \times M^N)$ matrix, as the dimension of the Hilbert space of the $N$-photon states in $M$ grids is $O(M^N)$. This can be reduced if the time evolution operator can be diagonalized but still requires manipulation of $O(M^N)$ elements.

To expedite the calculation, we exploit the fact that $N$-partite states such as $\ket{\Phi(t)}$ can be expressed with a fewer number of terms with a suitable choice of local states~\cite{schmidt}.
In addition our simulation only use the atoms to mimic linear optical objects such as mirrors and beamsplitters. 
These optical objects only need to reproduce their appropriate output electromagnetic wave for a given incoming wave. The atoms can effectively induce interactions between photons, but linear optical elements do not exhibit such photon-photon interaction. 
Based on this intuition, we simulate the time evolution of each photon separately neglecting the presence of the other photons in doing so.
More precisely, we consider $N$ atoms virtually located at the same position $\br_j$, which we call virtual atoms, and let the $N$ atoms only interact with their respective partner photon to simulate $N$-photon states.
This treatment limits interaction $\hat{h}_I$ within each single-photon system.
Our $N$-photon state becomes
\begin{eqnarray}
	\ket{\Phi(t)} &=&  \sum_{\xi_1, \cdots \xi_N} c_{\xi_1, \cdots \xi_N} 
 \ket{\phi_{\xi_1}(t)} \cdots \ket{\phi_{\xi_N}(t)} \label{eq:multi-photon} \\
 \ket{\phi_\xi(t)} &=&   \sum_{\br} c_\xi(t,\br) \ket{1_{\br}} + \sum_{j=1}^{N_A} c_{\xi, j}(t) \ket{1_{j}}, \nm
\end{eqnarray} 
where $\xi_n$ is a sets of the properties of $n$-th photon.
Deviation of this treatment from the true $N$-photon evolution occurs when two photons are present at the same location and an atom located at that point absorbs the photons. In our simulation, photon densities are kept sufficiently low to suppress this error.
These tricks render a time evolution operator of the $N$-photon system to be a direct product form as
\begin{eqnarray}
\hat{U}(t) = \exp(- \img \hat{h} t)  \otimes \cdots \otimes \exp(- \img \hat{h} t), \nm 
\end{eqnarray}
and the time evolution of the multi-photon system is written as,
\begin{align}
\ket{\Phi(t)} &= \hat{U}(t) \ket{\Phi(0)} \nm \\
&= \sum_{\xi_1, \cdots \xi_N} c_{\xi_1, \cdots \xi_N}
\left[\exp(- \img \hat{h} t)\ket{\phi_{\xi_1}(0)}\right] \cdots \left[\exp(- \img \hat{h} t)\ket{\phi_{\xi_N}(t)}\right] \nm \\
&= \sum_{\xi_1, \cdots \xi_N} c_{\xi_1, \cdots \xi_N} 
\ket{\phi_{\xi_1}(t)} \cdots \ket{\phi_{\xi_N}(t)} \nm.
\end{align}
Note that the dimension of each $\exp(- \img \hat{h} t)$ that needs to be compute is $M$, whereas the naive approach treats the time evolution operator $\hat{U}(t)$ of size $M^{N}$.
This reduction of the dimensionality dramatically improves speed of the time evolution and makes the multi-photon simulation feasible.

\subsection{Initial states}
All the initial states of the atoms at $t=0$ are set in their ground states.
For example, $c_{j}(0)=0$ in the case of the single-photon system, and
a Gaussian-shaped photon is injected in the space as
\begin{eqnarray}
c(0, \bk) =  \frac{2 \sigma \sqrt{\pi}}{L} 
\exp \left( -\frac{\sigma^2}{2}  (\bk -\bar{\bk})^2 -\img\bk \cdot \bar{{\bf r}} \right), \label{eq:single-photon_init}
\end{eqnarray}
where $\sigma$ is a width of the Gaussian. 
The vectors $\bar{\bk}$ and $\bar{\br}$ are parameters to decide the initial velocity and position, respectively.
For the simulations with the polarization, the initial state is given by 
\begin{eqnarray}
c_p(0,\bk) =  (\delta_{p, H} \cos \theta  + \delta_{p, V} \sin \theta) c(0, \bk), \nm 
\end{eqnarray}
where  $\theta$ is an initial polarization angle and  $\delta$ is the Kronecker delta.

As in Sec. \ref{subsec:hom} and Sec. \ref{subsec:chsh}, the two Gaussian-shaped photons are injected at $t=0$.
The modes of the photons are characterized by $\xi$ and $\eta$, where  $\xi$ and $\eta$ indicate sets of the photon
 properties that are $\sigma$, $\bar{\bk}$ and $\bar{\br}$.
For the HOM interference, 
we used an initial state as 
\begin{eqnarray}
\ket{\Phi(0)}=
\frac{1}{\sqrt{2}} \ket{\phi_\xi (0)}\ket{\phi_\eta (0)}
+ (\xi \leftrightarrow \eta), \nm
\end{eqnarray}
where $(\xi \leftrightarrow \eta)$ refers to the exchange of the indices in the previous term, which originates from the commutative relation of bosons.
For the test of the Bell-CHSH inequality, 
the initial state entangles their polarization $p$.
\begin{eqnarray}
 \ket{\Phi(0)}  &=& 
 \frac{1}{\sqrt{4}} \Bigl\{ \ket{\phi_{H,\xi} (0)} \ket{\phi_{H,\eta} (0)} 
 + \ket{\phi_{V,\xi} (0)} \ket{\phi_{V,\eta} (0)} \Bigr\} \nm \\
 &+&  (\xi \leftrightarrow \eta). \nm
\end{eqnarray}

\subsection{Computational details}

Table~\ref{tb:param_list} summarizes the parameters list that are used in the calculation shown in the main text.
All the simulations are performed by the time step $\delta t = 0.1$ and Gaussian width $\sigma = 2.0$.
Because we used the arbitrary unit by using $\hbar =c=1$, here we evaluate the spatio temporal scale in the real units. To recover the units,
at a typical optical wavelength of $500$ nm, the system size of the Mach-Zehnder interferometer (see Fig.~\ref{fig:mz}) is $\sim 0.025$ mm and simulation time $t = 46$ is corresponding to $0.12$ ps. Naively, for the lab scale simulation with a photon at an optical wavelength, we need to set a $10^5$ times larger space size.
To reduce the computational cost, Krylov subspace techniques~\cite{Branes2019, Michel2022} and quantics tensor trains~\cite{Shinaoka2023} might be applied.

\begin{table}[h]
  \centering
    \caption{Parameter list of the numerical calculation. The row and column names indicate corresponding figure names and parameters, respectively. The parameters $\bar{\br}$ and $\bar{\bk}$ are shown in Eq.~\ref{eq:single-photon_init}. The sixth column show the parameter of optical objects, such as mirror (M), beamsplitter (BS), phase shifter (PS), scatterer, polarization beamsplitter (PBS), and polarization rotator (PR). The parameters $D_j$, $\omega_j$, and $D_{j,p}$ are shown in Eqs.~\ref{eq:g_jk} and \ref{eq:g_jk_pol}.
  }
  \begin{tabular}{l|c|c|c|c|l}
    & Space size & Grid & $\bar{\br}$ & $\bar{\bk}$ & Optical components \\ \hline\hline
    \begin{tabular}{l}
    Fig.~\ref{fig:mz}\\
    (MZ)
    \end{tabular} & ($10\pi$, $10\pi$) & (256, 256) & (1.7, 7.7) & (10.0, 0) &
    \begin{tabular}{l}
    M: 88 atoms $\times$ 8 layers, $D_j=0.56$, $\omega_j=5.0$ \\
    BS: 88 atoms $\times$ 1 layers, $D_j=2.8$, $\omega_j=0.31$ \\
    PS: 120 atoms $\times$ 0$\sim$20 layers, $D_j=0.56$, $\omega_j=1.0$
    \end{tabular} \\ \hline
    \begin{tabular}{l}
    Fig.~\ref{fig:err_rate}\\
    (Scatterer)
    \end{tabular} & ($20\pi$, $10\pi$) & (512, 256) & (2.0, $5\pi$) & (10.0, 0) &
    Scatterer: 4 atoms $\times$ 4 layers, $D_j=1.0$, $\omega_j=5.0$ \\ \hline
    \begin{tabular}{l}
    Fig.~\ref{fig:HOM}\\
    (HOM)
    \end{tabular}  & ($15\pi$, $15\pi$) & (384, 384) &
    \begin{tabular}{c}
    $\xi$: (5.0, $7.5\pi$)\\
    $\eta$: ($7.5\pi$, 5.0)
    \end{tabular} & 
    \begin{tabular}{c}
    $\xi$: (5.0, 0)\\
    $\eta$: (0, 5.0)
    \end{tabular} &
    BS: 128 atoms $\times$ 1 layers, $D_j=2.0$, $\omega_j=0.34$ \\ \hline
    \begin{tabular}{l}
    Fig.~\ref{fig:chsh}\\
    (Bell-CHSH)
    \end{tabular}  & ($20\pi$, $10\pi$) & (512, 256) &
    \begin{tabular}{c}
    $\xi$: ($10\pi$, $5\pi$)\\
    $\eta$: ($10\pi$, $5\pi$)
    \end{tabular} & 
    \begin{tabular}{c}
    $\xi$: (-10.0, 0)\\
    $\eta$: (0, 10.0)
    \end{tabular} &
    \begin{tabular}{l}
    PBS: 148 atoms $\times$ 8 layers, $D_{j,V}=0.56$, $\omega_{j}=5.0$ \\
    PR: 128 atoms $\times$ 16 layers, $D_{j,S}=0.56$, $\omega_{j}=1.2$
    \end{tabular} \\ \hline
  \end{tabular}
  \label{tb:param_list}
\end{table}

\section*{Contributions}
All authors designed this research project and discussed the results.
J.O. and S.K. developed the code and performed the calculations.
A.S. provided theoretical analysis in the simulation results.

\section*{Competing interests}
The authors declare no competing interests.

\section*{Data availability}
The data are available from the corresponding author upon a reasonable request.

\section*{Code availability}
The code is downloadable at \url{https://github.com/ToyotaCRDL/QOsimulator}.

\section*{Supplementary information}
\subsection*{Hamiltonian of single-photon system}
In this section, we transform a Hamiltonian of the one-photon system proposed in a previous study \cite{Havukainen1999} in a form suitable for the main text.
According to Ref. \cite{Havukainen1999}, the one-photon Hamiltonian that involves $N_A$ two-level atoms is defined by
\begin{eqnarray}
\hat{h} &=& \hat{h}_F + \hat{h}_A + \hat{h}_I \nm \\ 
\hat{h}_F &=&  \sum_{\bk} \omega_{\bk} \hat{a}_{\bk}^{\dagger} \hat{a}_{\bk} \nm \\
\hat{h}_A &=&  \sum_{j=1}^{N_A} \omega_j \hat{\sigma}_{j}.  \nm
\end{eqnarray} 
The Hamiltonian of the free propagation of the photon, $\hat{h}_F$, consists of an annihilation operator $\hat{a}_{\bk}$ for one photon that has a wave-number vector $\bk$.
That of the two-level atoms, $\hat{h}_A$, is composed of the Pauli z operator $\hat{\sigma}_j$ of the $j$-th two-level atom.
The parameters $\omega_{\bk}$ and $\omega_j$ indicate eigen energies of the photon and two-level atom, respectively.
The $\hat{h}_I$ represents interaction between the photon and two-level atoms.

A quantum state of the corresponding system is written by
\begin{eqnarray}
\ket{\phi} &=& \sum_{\bk} c(\bk) \ket{1_\bk} + \sum_{j=1}^{N_A} c_{j} \ket{1_j}, \label{eq:phi_kbase}
\end{eqnarray} 
where $c(\bk)$ and $c_{j}$ are coefficients of the probability amplitudes.
 When operating $\hat{h}_A$ to the state, we obtain
\begin{eqnarray}
\hat{h}_A \ket{\phi} &=& - \sum_j \omega_j  \sum_{\bk} c(\bk)\ket{1_\bk} 
+ \sum_{j, j'} (\delta_{j,j'} -  \delta_{j\neq j'})\omega_{j}  c_{j'}\ket{ 1_{j'}} \nm \\
&=& - N_A \omega \sum_{\bk} c(\bk)\ket{1_\bk} 
+ \sum_{j, j'} (2\delta_{j,j'} -  1)\omega_{j} c_{j'}\ket{ 1_{j'}}  \nm \\
&=& - N_A \omega \sum_{\bk} c(\bk)\ket{1_\bk} 
+ 2 \sum_{j} c_{j} \omega_{j} \ket{ 1_{j}} 
- N_A \omega \sum_{j} c_{j} \ket{ 1_{j'}} \nm \\
&=& - N_A \omega \ket{\phi} +  2 \sum_{j} c_{j} \omega_{j} \ket{ 1_{j}} \nm \\
&\equiv& (-N_A \omega + \hat{h}_a) \ket{\phi}, \nm
\end{eqnarray}
where $\omega = \sum_j \omega_j / N_A$, and we defined $\hat{h}_a$ as
\begin{eqnarray}
\hat{h}_a =  \sum_{j=1}^{N_A} 2 \omega_j \hat{a}_{j}^{\dagger} \hat{a}_{j},  \label{eq:ha}
\end{eqnarray}
Therefore, the Hamiltonian becomes
\begin{eqnarray}
\hat{h} &=& \hat{h}_0 + \hat{h}_I -N_A \omega  \nm
\end{eqnarray} 
where $ \hat{h}_0  = \hat{h}_F + \hat{h}_a$.
Because the constant energy $-N_A \omega$ does not influence the time evolution, we omit this term when displaying the Hamiltonian in the main text.

The  Hamiltonian $\hat{h}_I$ that describes the dipole-dipole interaction between the two-level atoms and photons as
\begin{eqnarray}
\hat{h}_I &=& \sum_{j, \bk} (
    g(j, \bk)   \hat{a}_{j}^{\dagger}  \hat{a}_{{\bk}} 
  + g^*(j, \bk)   \hat{a}_{{\bk}}^{\dagger}  \hat{a}_{j}
) \\
g(j, \bk)  &=& -\frac{\img}{2L} \sqrt{\omega_\bk} D_j e^{\img \bk \cdot \br_j}. \nm 
\end{eqnarray} 
We set the frequencies of the two-level atoms and photon to be in resonance in simulation, thus $g(j, \bk)$ 
is approximated as
\begin{eqnarray}
g(j, \bk)  = -\frac{\img}{\sqrt{2}L} \sqrt{\omega_j} D_j e^{\img \bk \cdot \br_j}.
\end{eqnarray}




\end{document}